\documentstyle[12pt]{article}
\begin{document}
\centerline{\Large\bf Addendum to `Gravitational Geons in 1+1 Dimensions'}
\vskip .7in
\centerline{Dan N. Vollick}
\vskip .1in
\centerline{Irving K. Barber School of Arts and Sciences}
\centerline{University of British Columbia Okanagan}
\centerline{3333 University Way}
\centerline{Kelowna, B.C.}
\centerline{Canada}
\centerline{V1V 1V7}
\vskip .9in
\centerline{\bf\large Abstract}
\vskip 0.5in
\noindent
In a recent paper \cite{Vo1} I found gravitational geons in two classes of 1+1 dimensional theories of gravity.
In this paper I examine these theories, with the possibility of a cosmological constant, and find strong field gravitational geons.
In the spacetimes in \cite{Vo1} a test particle that
is reflected from the origin suffers a discontinuity in $d^2t/d\tau^2$. The geons found in this paper do not suffer from this problem.
\newpage
In a recent paper \cite{Vo1} I examined geons in 1+1 dimensional theories of gravity with a spacetime metric given by
\begin{equation}
ds^2=-f(r)dt^2+f(r)^{-1}dr^2\; ,
\end{equation}
where $r$ is a radial-like coordinate with $r\geq 0$. Imposing a reflecting boundary condition at $r=0$
leads to the problem that $d^2t/d\tau^2$ suffers a jump discontinuity upon reflection unless $f^{'}(0)=0$.
This follows from the geodesic equation
\begin{equation}
\frac{d^2t}{d\tau^2}=-\left(\frac{f'}{f}\right)\frac{dr}{d\tau}\frac{dt}{d\tau}
\end{equation}
with $dr/d\tau$ changing sign upon reflection. None of
the solutions in \cite{Vo1} satisfied this condition, except (13) with $B=0$. In this paper I find
additional strong field geon solutions
that either satisfy $f^{'}(0)=0$ or that can be extended to negative $r$ (i.e. with $-\infty<r<\infty$).

Consider the field equation (see equation (3) in \cite{Vo1} with $\alpha=0$)
\begin{equation}
R+\beta \Box R=\Lambda\; .
\end{equation}
In terms of the metric function $f(r)$ this equation can be written as
\begin{equation}
f^{''}+\beta\frac{d}{dr}\left(ff^{'''}\right)=-\Lambda\;.
\end{equation}
Integrating twice gives
\begin{equation}
f+\beta \left[ff^{''}-\frac{1}{2}\left(f^{'}\right)^2\right]=C_1+C_2r-\frac{1}{2}\Lambda r^2\;,
\label{eqn}
\end{equation}
where $C_1$ and $C_2$ are integration constants. In terms of $h=\sqrt{f}$ ($f>0)$ the above can
be written as
\begin{equation}
h^2+2\beta h^3h^{''}=C_1+C_2r-\frac{1}{2}\Lambda r^2\; .
\end{equation}
First consider the case $C_2=\Lambda=0$. Integrating once gives
\begin{equation}
\frac{1}{2}v^2+V(h)=E
\end{equation}
where $v=dh/dr$, $E$ is an integration constant and
\begin{equation}
V(h)=\frac{1}{2\beta}\left[\ln|h|+\frac{C_1}{2h^2}\right]\;.
\end{equation}
From this equation it is easy to see that $R=-f^{''}\simeq\frac{1}{\beta}\ln f$ for large $f$ (the solution
will approach a large vale of $f$ only if $\beta<0$). Thus,
we are only interested in solutions in which $f$ remains finite. The only nontrivial solutions with
$h^{'}(0)=0$ and which satisfy $h>0$ occur when $\beta$ and $C_1$ are positive. In this case
$V$ has one local minimum and goes to $\infty$ as $h$ goes to zero and $\infty$. Therefore
$f$ will undergo periodic oscillations and these solutions are similar to the solutions (13) in \cite{Vo1}, except that
they are not restricted to weak fields. Note that we can extend these solutions to negative values of $r$. These
solutions therefore correspond to infinite sequences of geons.

Now consider the case with $C_1=C_2=0$. In this case $f=-\frac{1}{2}\Lambda r^2$
($h=\sqrt{\frac{1}{2}|\Lambda|}r$) is a solution if $\Lambda<0$, so that $f>0$.
The behavior  of the solutions can be analyzed by letting $h=rZ(r)$ and $x=\ln|r|$.
The field equation in these variables is given by
\begin{equation}
\frac{d^2Z}{dx^2}=-\frac{dZ}{dx}-\frac{dU(Z)}{dZ}\;,
\end{equation}
where the potential is given by
\begin{equation}
U(Z)=\frac{1}{2\beta}\left[\ln|Z|-\frac{\Lambda}{4Z^2}\right]\;.
\end{equation}
For $\Lambda <0$ and $\beta>0$ then $U\rightarrow \infty$ as $Z\rightarrow
0,\infty$ and there is one minimum at $Z=\sqrt{|\Lambda|/2}$. The damping term $-dZ/dx$ will cause the motion to
decay to the solution $Z=\sqrt{|\Lambda|/2}$ at large $r$. Thus, at large $r$ the function $f(r)$ will approach $\frac{1}{2}|\Lambda|r^2$.
Note that this solution can be extended to negative values of $r$ since $Z\rightarrow\sqrt{|\Lambda|/2}$ for large
$x$ where $x=\ln|r|$. This solution therefore corresponds to a geon with $-\infty<r<\infty$ or with $r\geq 0$ and a reflective
boundary condition at $r=0$ ($f'=2rZ^2+2r^2ZZ'$, so that $f'(0)=0$ if
$Z(0)$ and $Z'(0)$ are finite). At large $r$ the Ricci scalar approaches $-\Lambda$. There are no geon solutions for $\beta<0$.

Now consider the case $C_1=\Lambda=0$. A solution of this equation is $f=C_2r+\frac{1}{2}\beta C_2^2$. I have been
unable to show analytically that, for $\beta>0$ and $C_2>0$, all solutions approach this solution for large $r$. However, it is easy to show
that this solution is stable in the sense that all nearby solutions do converge to it for large $r$. For simplicity I will take $C_2=1$. Now Let
\begin{equation}
f(r)=\left[r+\frac{1}{2}\beta \right]\left(1+g(r)\right)
\label{g}
\end{equation}
The function $g(r)$ satisfies the linearized equation
\begin{equation}
\beta\left[x^2\frac{d^2g}{dx^2}+x\frac{dg}{dx}-g\right]+xg=0\;,
\end{equation}
where $x=r+\beta$. Now define $y=2\sigma\sqrt{x}$,
where $\sigma=\beta^{-1/2}$. In terms of $y$ the equation is
\begin{equation}
y^2\frac{d^2g}{dy^2}+y\frac{dg}{dy}+(y^2-4)g=0\;.
\end{equation}
This is a Bessel equation of order two, so the general solution is
\begin{equation}
g(x)=AJ_2\left(2\sigma\sqrt{x}\right)+BY_2\left(2\sigma\sqrt{x}\right)\;,
\end{equation}
where $A$ and $B$ are constants. Here I have included $Y_2$, which diverges at the origin, since I am considering
only large $x$. Note that
\begin{equation}
J_2(x)\simeq\sqrt{\frac{2}{\pi x}}\cos\left(x-\frac{5}{4}\pi\right)\;\;\;\;\; as\;\;\;\;\; x\rightarrow\infty
\end{equation}
(a similar expression holds for $Y_2$). Thus, $g$ goes to zero for large $r$ and $f\rightarrow r+\frac{1}{2}\beta$ for
large $r$. In this case $R=-f^{''}$ goes to zero at large $r$. I have studied the differential equation (\ref{eqn}) with (\ref{g})
numerically ($C_1=0,C_2=1,\Lambda=0$),
using Maple, for the initial values $g(0)\in[-0.9,3]$ and with $\beta=1$. The value of $g'(0)$ is determined by setting $f'(0)=0$. This
gives $g'(0)=-2[1+g(0)]$.
The solutions oscillate and decay slowly
for large $r$ in a similar fashion to the linearized solutions.

The second class of theories examined in \cite{Vo1} was based on the Lagrangian
\begin{equation}
L=-\sqrt{g}\left[\frac{1}{\phi}R+V(\phi)\right]\;.
\end{equation}
It was shown that the field equations could be integrated giving
\begin{equation}
Af^{'}=V(\phi) \;\;\;\;\;\;\;\; with\;\;\;\;\;\;\; \phi=\frac{1}{Ar}\;,
\end{equation}
where $A$ is a constant. Thus, for a given $f(r)$ it is easy to solve for $V(\phi)$.
A simple function that satisfies $f^{'}(0)=0$ and has Schwarzschild behavior at large $r$ is
\begin{equation}
f(r)=1-\frac{2mr^2}{r^3+2m\ell^2}\; ,
\end{equation}
where $m$ and $\ell$ are constants. The potential is given by
\begin{equation}
V(\phi)=\frac{2mA^2\phi^2\left(1-4m\ell^2A^3\phi^3\right)}{\left(1+2m\ell^2A^3\phi^3\right)}\;.
\end{equation}
\section*{Acknowledgements}
This research was supported by the  Natural Sciences and Engineering Research
Council of Canada.


\begin{thebibliography}{99}
\bibitem{Vo1}
Dan N. Vollick Class. Quant. Grav. 25, 175004 (2008)
\end{thebibliography}
\end{document}